\documentclass[fleqn,twoside]{article}
\usepackage{espcrc2}

\usepackage{graphicx}
\usepackage{epsfig}

\newcommand{\bel}[1]{\begin{equation}\label{#1}}
\newcommand{\bal}[1]{\begin{eqnarray}\label{#1}}
\newcommand{\ee}{\end{equation}}
\newcommand{\ea}{\end{eqnarray}}
\newcommand{\equ}[1]{~Eq.(\ref{#1})}

\newcommand{\bmath}{\begin{displaymath}}
\newcommand{\emath}{\end{displaymath}}
\newcommand{\bite}{\begin{itemize}}
\newcommand{\eite}{\end{itemize}}

\newcommand{\bx}{{\bf x}}

\newcommand{\tr}{{\rm Tr}}

\newcommand{\drop}[1]{}

\newcommand{\AmS}{{\protect\the\textfont2
 A\kern-.1667em\lower.5ex\hbox{M}\kern-.125emS}}

\title{Confinement at Weak Coupling}

\author{Martin Schaden\thanks{email: mschaden@andromeda.rutgers.edu}\\
{Rutgers University, 211 Smith Hall, 101 Warren St., Newark, NJ 07102, U.S.A.}%
        }

\begin{document}

\begin{abstract}
The free energy of $U(N)$ and $SU(N)$ gauge theory was recently
found to be of order $N^0$ to all orders of a perturbative
expansion about a center-symmetric orbit of vanishing curvature.
Here I consider extended models for which this expansion is
perturbatively stable. The extreme case of an $SU(2)$ gauge theory
whose configuration space is restricted to center-symmetric orbits
has recently been investigated on the lattice~\cite{Langfeld05}.
In extension of my talk, a discussion and possible interpretation
of the observed finite temperature phase transition is given. The
transfer matrix of constrained $SU(N)$ lattice gauge theory is
constructed for any finite temperature.\vspace{1pc}
\end{abstract}

\maketitle

\section{INTRODUCTION}
\label{Intro} The non-perturbative dynamics of $SU(N)$ gauge
theories simplifies considerably for a large number of colors
$N$~\cite{Makeenko04} and confinement can be seen in a
perturbative expansion of the free energy about a center-symmetric
ground state configuration~\cite{Schaden05}. The center-symmetric
perturbative vacuum has the same classical action as the usual
perturbative vacuum for which the center symmetry is broken.
However, the 1-loop free energy indicates that $SU(N)$ is in the
broken phase~\cite{Weiss81} at weak coupling. The center symmetry
thus is broken at sufficiently high temperatures and a
perturbative center-symmetric expansion is somewhat formal.
However, this (formal) perturbative expansion to all orders in the
coupling defines a center-symmetric $1/N$-expansion. I here
consider some modified models whose center-symmetric phase is
stable at any finite order of a weak coupling expansion. To mimic
non-perturbative and higher order perturbative contributions to
the free energy, external sources that couple to adjoint Polyakov
loops are included~\cite{Meisinger97}. Although such sources
probably cannot be physically realized~\cite{Semenoff97}, the
approach is analogous to studying a spin model in a magnetic
field: the external sources favor and stabilize the weak coupling
expansion about a particular (in this case center-symmetric)
classical ground state.

The primary objective here is to construct models that confine at
weak coupling. Only minimal modifications to the standard Wilson
action are considered that may achieve this. There in particular
is no attempt to construct a general and realistic effective
action for Polyakov loops~\cite{loopactions02}.

In the limit of very large external sources, the configuration
space consists of center-symmetric configurations only.
Constraining the configuration space of $SU(2)$-LGT to
center-symmetric orbits only, was recently~\cite{Langfeld05} found
to nevertheless reproduce the full non-perturbative lattice
$\beta$-function for $T\rightarrow 0$. Somewhat surprisingly, this
constrained $SU(2)$-model shows a violent increase in entropy at
roughly the usual deconfinement temperature even though
center-symmetry is enforced~\cite{Langfeld05}.

\section{CENTER SYMMETRIC ORBITS}
Center symmetric orbits are determined by their Polyakov loops.
Consider a particular configuration of links $\{U\}=\{U_{x\mu}:
U_{x\mu}\in SU(N),\ \forall x=(\bx,\tau), \mu=1,\dots, 4\}$ of a
4-dimensional periodic lattice with periods $N_T$ and $N_s$ on
temporal and spatial cycles respectively. The temporal Polyakov
loops, ${\cal P}^{(k)}(\bx)$, at a site $x=(\bx,\tau)$ are,
\bal{Ploop}
{\cal P}^{(k)}(\bx)&=&\tr U^k(x) \\
U(x)&=&U_{(\bx,\tau)4}U_{(\bx,\tau+1)4}\dots U_{(\bx,\tau+N_T)4}\
.
\ea
${\cal P}^{(k)}(\bx)$ does not depend on the Euclidean time $\tau$
and
\bel{Q}
Q^{(k)}[\{U\}]:=\sum_{\bx} {\cal P}^{(k)}(\bx){\cal
P}^{(k)\dagger}(\bx)\ge 0\ ,
\ee
may be thought of as gauge invariant topological charges of the
configuration. The $Q^{(k)}$ are non-negative by definition and
(up to a field-independent term) can be expressed by Polyakov
loops in the adjoint representation. Of particular interest is
$Q=Q^{(1)}$ given by ordinary Polyakov loops that wind just once
around the temporal direction.

A (global) center transformation with the center element $z {\bf
1}\in {\bf Z}_N\subset SU(N)$ maps $\{U\}$ to the configuration
$\{U\}_z$ (up to possibly a periodic gauge transformation),
\bal{ctrafo}
\{U\}_z=&&\{U^\prime_{(\bx, 1)4}=z U_{(\bx, 1)4},\forall \bx \
{\rm and}\nonumber\\&& \quad U^\prime_{x\mu}=U_{x\mu}\ {\rm
otherwise}\}
\ea
The Wilson plaquette actions of $\{U\}$ and $\{U\}_z$ are
identical. However, the center transformation of\equ{ctrafo}
\emph{in general} is not equivalent to a gauge transformation of
the configuration: gauge invariant Polyakov loops change by ${\cal
P}^{(k)}(\bx)\rightarrow z^k {\cal P}^{(k)}(\bx)$.

A configuration is center-symmetric, if any center transformation
can be undone by a periodic gauge transformation. The gauge orbit
represented by a center-symmetric configuration thus does not
change under center transformations and we have that,

\emph{center-symmetric $\{U\}$} $\Leftrightarrow
Q^{(1)}[\{U\}]=0$.

The definition of $Q=Q^{(1)}$ in\equ{Q} implies that \emph{all}
temporal Polyakov loops of a $Q=0$-configuration vanish. Since all
contractible Wilson loops also are not affected by the center
transformation, the orbit does not change. If on the other hand
$Q>0$, \emph{some} Polyakov loop ${\cal P}^{(1)}(\bx)\neq 0$. This
gauge-invariant quantity acquires a phase and changes under center
transformations. The orbit therefore changes when $Q[\{U\}]>0$.

Note that the non-local nature of the charge $Q$ in\equ{Q} is due
to its gauge invariant definition only. With a gauge
transformation temporal links on all but one particular temporal
slice of a periodic lattice can be set to unity. In such an axial
gauge, $Q$ is a quadratic function of the remaining non-trivial
temporal links only. Due to the periodicity of the lattice, a
gauge transformation that does not depend on the Euclidean time
\emph{diagonalizes} the remaining non-trivial temporal links and
\emph{orders} their eigenvalues. Finally, an Abelian gauge
transformation will evenly distribute the diagonal temporal links
in Euclidean time~\cite{Schaden05}.

Center-symmetric gauge orbits of a lattice gauge theory (LGT) with
SU(N) structure group can thus be represented by standardized
configurations with,
\bal{specs}
U_{x4}&=& G=e^{2\pi i\bar\theta/N_T},\ {\rm where}
\\ \bar\theta &=&{\rm diag}(\frac{N-1}{2N},\frac{N-3}{2N},\dots,\frac{1-N}{2N}) \
.\nonumber
\ea
The gauge transformation that brings a center-symmetric
configuration to the standard form of\equ{specs} furthermore is
unique up to Abelian gauge transformations that do not depend on
Euclidean time.  Since none of the eigenvalues of $U(x)$ are
degenerate when $Q[\{U\}]=0$, there is no monopole gauge fixing
ambiguity~\cite{tHooft74} for center-symmetric orbits. They
therefore do not describe monopole configurations. The results
of~\cite{Reinhardt97} imply that the Pontryagin index of
center-symmetric configurations vanishes as well.

Due to the complete determination of the temporal links in a
particular axial (Polyakov) gauge, the transfer matrix method can
be used to describe the dynamics of $SU(N)$ gauge theory
constrained to center-symmetric orbits. There are two main
differences to the transfer matrix for the space of gauge orbits
in axial gauges of~\cite{Creutz77}. Because the constraint to
center-symmetric orbits determines \emph{all} temporal links of a
periodic lattice uniquely in a particular gauge, one can formulate
the partition function in terms of a transfer matrix even for
periodic lattices of \emph{finite} extent, i.e. when $T<\infty$.
In addition, the residual gauge group is the group of
time-independent Abelian gauge transformations only and Gauss's
law for selecting physical states simplifies accordingly.

Identifying a configuration of spatial links $\{U_{(\bx,0)\mu}\}$
with a basis vector $|U\rangle$ of the Hilbert space, the transfer
matrix elements of the constrained center-symmetric Yang-Mills
theory are,
\bal{transf}
\langle U^\prime|T|U\rangle &=& e^{{\rm Re}\tr
\Phi(\{U^\prime,U\})}\, ,
{\rm with}\\
\Phi(\{U^\prime,U\})&&\hspace{-2.5em}=\beta_T\sum_{\bx,i}(U^{\prime\dagger}_{\bx
i}G^\dagger U_{\bx i}G ) +\beta_s \sum_{j\neq
i}P_{\bx,ij}\nonumber
\ea
where $P_{\bx,ij}$ is the plaquette in the spatial $ij$-plane at
$\bx$. $\beta_T$ and $\beta_s$ are inverse coupling constants
related to the temporal and spatial lattice spacing by dimensional
transmutation. The dependence of the kinetic term on the diagonal
matrix $G$ given in\equ{specs} ensures that only center-symmetric
configurations are generated by the transfer matrix
of\equ{transf}. Note that the parallel transport by $G$ depends on
$N_T$. For fixed lattice spacing, $G\sim {\bf 1}$ at sufficiently
low temperature and the transfer matrix in\equ{transf} essentially
becomes the one of Creutz~\cite{Creutz77}. One is tempted to argue
that the constraint to center-symmetric orbits is irrelevant at
low temperatures and that the full $SU(N)$ model should therefore
be recovered.  But the residual gauge symmetry of the
center-symmetric transfer matrix at any finite temperature $T>0$
is just the Cartan subgroup rather than the full $SU(N)$ gauge
group. The observation by~\cite{Langfeld05} that a simulation of
the constrained SU(2) lattice model reproduces the Wilson loop
expectation value of the full $SU(2)$-LGT thus is rather
encouraging.

Since all Polyakov loops with non-trivial N-ality vanish for a
center-symmetric configuration, their expectation values cannot be
used as order parameters of the constrained theory. The sudden
increase in entropy of constrained $SU(2)$ found
in~\cite{Langfeld05} thus is not the manifestation of a phase with
broken center symmetry, even though the transition occurs at
roughly the deconfinement temperature of $SU(2)$-LGT. The entropy
density of the constrained model in fact exceeds the
Stefan-Boltzmann limit for free massless gluons by an order of
magnitude and is rather sensitive to the spatial lattice volume.
This behavior of the truncated $SU(2)$-model could be explained by
a Hagedorn transition~\cite{Hagedorn65} at $T_H$. Instead of
deconfining color charges, the string tension vanishes at $T_H$
and a plethora of low-mass glueball states of arbitrary high spin
is produced. The entropy density for $T>T_H$ in this case is
limited by the infra-red cutoff of the finite lattice only.

\section{PERTURBATIVE STABILITY}

A Hagedorn transition at $T_H$ can perhaps be described by the
Nambu-Goto string model~\cite{Pisarski82} and may even allow a
perturbative expansion~\cite{LargeNHag04} of the constrained model
just below $T_H$ at large $N$. In support of weak coupling, the
Stefan-Boltzmann bound apparently is attained by the constrained
SU(2)-model \emph{just below or at} $T_H$ (see Fig.3b of
ref.~\cite{Langfeld05}).

To better understand the behavior of models that confine at weak
coupling, I will now consider the perturbative free energy of
$U(N)$ and $SU(N)$ models with extended lattice actions,
\bel{actions}
S(\beta,\vec\kappa)=\beta S_W[\{U\}]+ N_T
\sum_{k=1}^{[N/2]}\kappa_k Q^{(k)}[\{U\}] .
\ee
Here $S_W[\{U\}]$ is Wilson's plaquette action. The charges
$Q^{(k)}[\{U\}]$ are defined by\equ{Q} and the parameters
$\kappa_k\geq 0, k=1,\dots,[N/2]$ can be thought of as
thermodynamic potentials or external sources\footnote{$[N/2]$ is
the largest integer that satisfies $[N/2]\leq N/2$.}.

To retain center symmetry, the Wilson action is extended by
center-symmetric terms only. Note that Polyakov loops in the
adjoint representation renormalize
multiplicatively~\cite{Makeenko04} and scale like the Wilson
action at large $N$. Up to a field-independent normalization
constant, the divergence may be absorbed in the parameters
$\kappa_k$.

$\kappa_1>0$ emphasizes center-symmetric configurations and the
extended models should therefore approach one with
center-symmetric orbits only~\cite{Langfeld05} for
$\kappa_1\sim\infty$. For $SU(N>3)$ additional sources for charges
$Q^{(2)}\dots Q^{([N/2])}$ are needed to have a well-defined weak
coupling expansion in the center-symmetric phase. I include only
the minimal number of external sources that stabilize the
center-symmetric perturbative ground state.

The non-Wilson terms of the extended actions of\equ{actions} lift
the degeneracy of the lattice ground state with respect to the
center symmetry. $Q=Q^{(1)}[\{U\}] \geq 0$ breaks the degeneracy
completely and for $\kappa_k>0,k=1,\dots,[N/2]$, configurations
with minimal classical action are on center-symmetric orbits
without curvature~\cite{Schaden05}. This center-symmetric orbit
with lowest classical action furthermore is unique for a spatial
lattice topology\footnote{For instance the lattice equivalent of
$S_3\times S_1$. The spatial part of this lattice is the
3-dimensional surface of a 4-dimensional hyper-cube. Its 16
"corners" have (spatial) connection number five instead of six as
for all other sites.} with contractible\footnote{On the lattice
closed loops of links are considered similar when they differ in
two consecutive links only. Two lattice loops are homotopic if a
series of such similar loops interpolates between them. Finally, a
lattice loop is contractible if it is homotopic to the "loop"
without links. Plaquettes in this sense are contractible and
Polyakov loops are not contractible on a periodic lattice.}loops
only.

$\kappa_1>0$ with $\kappa_i=0$ for $i>1$ is sufficient for a
unique center-symmetric classical vacuum but does not stabilize
the weak coupling expansion about this vacuum for $N>3$. Although
the minimum of $Q^{(1)}[\{U\}]$ occurs at the center-symmetric
configurations, it is at least of third order in $N-3$ directions.
This can be seen by expanding $Q^{(k)}$ to second order in
fluctuations about a center-symmetric configuration in the
standard form of\equ{specs}. Due to gauge invariance it is
sufficient to consider Abelian fluctuations of the temporal links
that do not depend on Euclidean time,
\bel{fluct}
Q^{(k)}[\{\theta^i-\bar\theta^i=\delta^i\}]=\sum_{{{\rm
spatial}\atop{{\rm sites}\ i}}}\sum_{a,b=1}^N \delta_a^i
H^{(k)}_{a b}\delta_b^i+\dots\,
\ee
with $N\times N$ Hessians,
\bel{Hessians}
H^{(k)}_{ab}= \cos(2\pi k(a-b)/N)\ ,
\ee
that are diagonal in the Fourier basis,
\bel{Fourier}
\theta^{(s)}_a=e^{2\pi i s a/N},\ j=1,\dots,N.
\ee
$\theta^{(N)}=(1,\dots,1)$ changes the determinant of the links
and is a zero-mode of all $H^{(k)}$. $SU(N)$ fluctuations are in
the $N-1$-dimensional sub-space spanned by $\{\theta^{(s)};0<s<
N\}$. For $N>3$ the minimum of $Q=Q^{(1)}$ is of third and higher
order in directions spanned by the eigenvectors $\theta^{(s)}$
with $s=2,\dots,N-2$. The Hessian $H^{(1)}$ thus has additional
zero-modes when $N>3$. To obtain a quadratic minimum suitable for
a perturbative expansion about the center-symmetric configuration,
the Wilson action should be extended by charges $Q^{(k)}$ with
$k>1$ for $SU(N>3)$. For $SU(2)$ and $SU(3)$ lattice groups, an
extension by $\kappa_1 Q^{(1)}[\{U\}]$ is sufficient.

\section{THE 1-LOOP FREE ENERGY}
Since it is gauge invariant, the 1-loop free energy is
obtained~\cite{Schaden05,Weiss81} by expanding around a background
of constant Abelian links,
\bel{Abelian}
\bar U_{x4}= {\rm diag}\;(e^{2\pi i\theta_1/N_T},\dots,e^{2\pi
i\theta_N/N_T}),
\ee
with $\bar U_{x1}=\bar U_{x2}=\bar U_{x3}={\bf 1}$. In the
thermodynamic- and continuum- limit the free energy density of the
extended $U(N)$ model at weak coupling is~\cite{Schaden05},
\bal{F0}
F_0(T,U(N))&=& 2(1-N_A)\sum_{a,b=1}^N
I(T,\theta_a-\theta_b;0)\nonumber\\&&\hspace{-2em}-
4\sum_{j=1}^{N_F}\sum_{a=1}^N
I(T,\theta_a+1/2;m_j)\\&&\hspace{-2em}+\sum_{k=1}^{[N/2]}\tilde\kappa_k\sum_{a,b=1}^N\cos[2\pi
k(\theta_a-\theta_b)]\ .\nonumber
\ea
$N_A$ here is the number of Majorana fields in the adjoint
representation satisfying \emph{periodic} boundary
conditions\footnote{$N_A=1$ and $N_F=0$ is (${\cal N}=1$)
supersymmetric Yang-Mills (SYM) and the Majorana particle in this
case is the gaugino. Only $N_A=0$ was considered
in~\cite{Schaden05}.} and $N_F$ is the number of
\emph{anti-periodic} Dirac fields in the fundamental
representation. The constants $\tilde\kappa_j$ have mass dimension
four and are the (renormalized) continuum analogs of the
$\kappa_j$ in the lattice model.  The function $I(T,\delta;m)$
is~\cite{Schaden05},
\bel{I}
I(T,\delta;m)=-T^2 m^2\sum_{n=1}^\infty \frac{\cos(2\pi
n\delta)}{2\pi^2 n^2} K_2({\frac{n m} T}),\\
\ee
where $K_2$ is a K-Bessel function normalized so that $K_2(|z|\sim
0)=2/z^2$. In the limit of vanishing mass \equ{I} therefore
simplifies to,
\bel{I0}
I(T,\delta;0)=-
T^4\sum_{n=1}^\infty \frac{\cos(2\pi n\delta)}{\pi^2 n^4}.
\ee

The one-loop free energy of the extended $SU(N)$ model is related
to that of the extended $U(N)$ model by noting that the
$U(1)$-photon as well as one of the of the Majorana fields
essentially decouple. To one loop one therefore has,
\bel{F0SU}
F_0(T,SU(N))= F_0(T,U(N))+\frac{\pi^2(1-N_A) T^4}{45}\ .
\ee
The correction term is of order $N^0$ and does not depend on the
angles $\theta_a$.  However, the decoupled U(1)-"photon" and
Majorana fields contribute to the pressure and entropy density of
the $U(N)$-model only. This contribution in principle could be
critical for \emph{thermodyamic} stability which requires a
positive entropy density and pressure. Since the decoupled fields
give a phase-independent contribution to the free energy, one can
ignore this physical requirement by assuming a contribution from
"inert" degrees of freedom that are irrelevant for confinement but
guarantee an overall positive pressure and entropy
density\footnote{For $N_A=0$ the "inert" degree of freedom could
for instance simply be the physical photon.} In the following I
therefore do not distinguish between the one-loop free energy of
$U(N)$- and $SU(N)$- gauge models.

\subsection{Perturbative Stability of the Free Energy}
It is straightforward to verify that the center-symmetric
configuration with $\theta_a=\bar\theta_a$ of\equ{specs} is an
extremum of the 1-loop free energy of\equ{F0}. It furthermore is
well known~\cite{Weiss81} that this extremum is a \emph{maximum}
of the perturbative free energy for vanishing $N_A$ and
$\tilde\kappa_k$'s.  The center-symmetric ground state in this
case is not stable at weak coupling and the conclusion is that
$SU(N)$ gauge theories are in a non-confining plasma phase at
sufficiently high temperatures~\cite{Weiss81}. When $N_A>1$ or the
$\tilde\kappa_k$ are sufficiently large, this need not be the
case.

Without Dirac fields, that is for $N_F=0$, the center-symmetric
configuration is \emph{perturbatively} stable~\cite{Hosotani89}
when $N_A>1$. Note that for supersymmetric Yang-Mills with
$N_A=1$, arbitrary small $\tilde\kappa_k>0$ are capable of
stabilizing the center-symmetric ground state at weak coupling.
The exact perturbative cancellation between the gluon- and
gaugino- contributions to the free energy is a manifestation of
the supersymmetry of the model when $N_A=1$. At finite temperature
this cancellation occurs only for a gaugino satisfying
\emph{periodic} boundary conditions and the supersymmetry is
(softly) broken by anti-periodic Majorana
fields~\cite{Takenaga98}.

In the absence of massless \emph{periodic} matter fields in the
adjoint representation, center-symmetric configurations minimize
the 1-loop free energy of $SU(N)$ at (sufficiently) low
temperatures when $\tilde\kappa_i>0$. To obtain an upper bound for
the critical temperature of these models, consider the Hessian of
the 1-loop free energy at the center-symmetric configuration
$\bar\theta_a$,
\bal{Hess}
{\cal H}_{ab}&:=&\left.\frac{1}{8\pi^2 T^2}\frac{\partial^2
F_0}{\partial\theta_a\partial\theta_b}\right|_{\theta=\bar\theta}\\
&&\hspace{-4em}= \frac{2T^2(1-N_A)}{\pi^2}\sum_{n=1}^\infty
\left[\frac{\delta_{ab}}{N n^2}-\frac{\cos\left(\frac{2\pi n
(a-b)}{N}\right)}{n^2}\right]\nonumber\\
&&\hspace{-4em}+\sum_{n=1}^{[N/2]}\frac{n^2\tilde\kappa_n}{T^2}
\cos\left(\frac{2\pi n(a-b)}{N}\right)\nonumber\\
&&\hspace{-4em}-\sum_{j=1}^{N_F}\frac{m_j^2}{\pi^2}\sum_{n=1}^\infty
\delta_{ab}\cos\left(\frac{(2 a-1)\pi n}{N}\right)K_2({\frac{n
m_j} T})\ .\nonumber
\ea
Without matter in the fundamental representation ($N_F=0$), ${\cal
H}_{ab}$ is diagonal in the Fourier basis of\equ{Fourier} with
eigenvalues $h^{(j)}$ for $j=1,\dots,N-1$ given by,
\bal{eigen}
h^{(j)}&=&\frac{T^2(1-N_A)}{6
N}\left(1-\frac{3}{\sin^2(j\pi/N)}\right)\nonumber\\&& +\frac{N}{2
T^2}(j^2\tilde\kappa_j+(N-j)^2\tilde\kappa_{N-j}),
\ea
where $\tilde\kappa_j=0$ for $j>[N/2]$ and $j<1$. A perturbative
expansion about the center-symmetric orbit of minimal classical
action requires that the Hessian is positive definite, or
$h^{(j)}>0,j=1,\dots,N-1$. For $N_A>1$ the center-symmetric model
is stable with respect to small perturbations for vanishing
$\vec{\tilde\kappa}=0$ and at \emph{all} temperatures $T>0$. Any
transition to a phase with broken center symmetry at $T_c<\infty$
in this case is of first order. For $N_A<1$ on the other hand, the
Hessian is positive for $T<T_p$ only, where
\bel{Tp}
T^4_p=\min_{j}\frac{N^2
(j^2\tilde\kappa_j+(N-j)^2\tilde\kappa_{N-j})}{2(1-N_A)(\sin^{-2}(j\pi/N)
-1/3)}\ .
\ee
The $\tilde\kappa$'s thus limit the temperature range for which a
weak coupling expansion about the center-symmetric background
could be stable. For $SU(2)$ and $SU(3)$ gauge groups there is
just one $\tilde\kappa$ to contend with and \equ{Tp} implies
perturbative stability of the purely gluonic models with
$N_A=N_F=0$ for $T<T_p$ where
\bel{Tp23}
T^4_{p\;SU(2)}=6\tilde\kappa_1\ ;\
T^4_{p\;SU(3)}=\frac{9}{2}\tilde\kappa_1\ .
\ee
Note that since $\tilde\kappa_1\rightarrow\infty$ essentially
constrains the configuration space to center-symmetric
orbits~\cite{Langfeld05}, the classical center-symmetric vacuum
apparently is perturbatively stable at all temperatures in this
limit. However, $T_p$ is only an upper bound and  a first order
Hagedorn transition may, and apparently does~\cite{Langfeld05},
occur at a finite temperature $T_H<\infty=T_p$ in $SU(N)$-LGT.

It perhaps is worth mentioning that $T_p$ becomes independent of
$N$ if $\tilde\kappa_j(N) \propto (N \sin(j\pi/N))^{-2}$ for large
$N$. The extended models thus are stable with respect to small
perturbations over a finite range of temperature with
$\tilde\kappa_j(N)$ of order $N^0$.

\subsection{The Phase Transition of the Extended U(2) and U(3) Models}
Perturbative stability is necessary, but not sufficient for
thermodynamic stability.  However, for $\tilde\kappa_j>0$, the
global minimum of the free energy of the extended $SU(N)$ models
at sufficiently small coupling is center symmetric for $T\sim 0$.
If this perturbatively stable state turns into a \emph{local}
minimum at a temperature $T_c<T_p$, a first order transition to
the true minimum of the free energy may occur. At sufficiently
weak coupling, the center symmetry then would be broken for
temperatures above $T_c$.

The free energy of the extended $U(2)$ model with no quark flavors
depends on just one angle $\theta=\theta_2-\theta_1$. The 1-loop
free energy is,
\bal{FSU2}
F_0(T,U(2))&=& -8(1-N_A)T^4 \sum_{n=1}^\infty \frac{\cos^2(\pi
n\theta)}{\pi^2
n^4}\nonumber\\&&+4\tilde\kappa_1\cos^2(\pi\theta)\ .
\ea
Its extrema are at half-integer values of $\theta$ and the
center-symmetric value of $\theta=1/2$ is the absolute minimum at
weak coupling for $T<T_c$, where
\bel{TcSU2}
T^4_{c\;SU(2)}=\frac{48}{\pi^2}\tilde\kappa_1<T^4_{p\;SU(2)}=6\tilde\kappa_1\
.
\ee
The first order transition of the extended $SU(2)$ model occurs
just below the temperature $T_p$ at which the center-symmetric
state anyway becomes unstable with respect to small fluctuations.
Assuming that the scale $\tilde\kappa_1$ does not change
drastically, the two temperatures are reasonably close,
$T_c=(8/\pi^2)\sim 0.95\; T_p$.

Note that universality arguments and lattice simulations suggest
that the deconfinement transition of SU(2)-LGT is of second
order~\cite{Yaffe82}. However, these arguments presume the
existence of a mass gap which does not arise at finite orders of
perturbation theory.

The analysis is more involved for $SU(N>2)$ and I only give
results for the extended $U(3)$ model. The free energy of $U(3)$
depends on two independent angles,
$\theta=\theta_1-\theta_2\leq\phi=\theta_1-\theta_3$. For $N_F=0$
and at 1-loop it is of the form,
\bal{F0SU3}
F_0(T,U(3))&=& \tilde\kappa_1\left|1+e^{2\pi i \theta}+e^{2\pi i
\phi}\right|^2\\&&\hspace{-4em}-2 T^4(1-N_A)\sum_{n=1}^\infty
\frac{\left|1+e^{2\pi i n \theta}+e^{2\pi i n
\phi}\right|^2}{\pi^2 n^4}\ .\nonumber
\ea
At weak coupling the center symmetry is broken in a \emph{first}
order transition at $T_{c_1\,SU(3)}\sim 0.92\; T_{p\,SU(3)}$. This
transition is rather similar to that of the $SU(2)$-model and in
fact is to a $U(2)$-invariant vacuum configuration. After the
transition, the Polyakov loop reaches $1/3$ of its maximal value
only. A subsequent first order transition to the perturbative
$SU(3)$-invariant vacuum occurs at $T_{c_2\,SU(3)}\sim 1.02
\;T_{p\,SU(3)}$. Again assuming that $\kappa_1(SU(3))$ does not
change appreciably, the two transitions are only about $10\%$
apart, that is $T_{c_2\,SU(3)}\sim 1.10 T_{c_1\, SU(3)}$.

\section{DISCUSSION}
At low temperatures all the $U(N)$ and $SU(N)$  models considered
here have a stable \emph{perturbative} ground state that is
invariant under the global $Z(N)$ center symmetry. The expectation
value of the Polyakov-loop vanishes to all orders in perturbation
theory in this low-temperature phase which "confines" static color
charges in this sense. However, there is no mass gap at any finite
order in the weak coupling expansion and all excitations are
massless. The effective number of thermodynamic degrees of freedom
in the perturbative center-symmetric phase nevertheless almost
vanishes: \equ{FSU2} implies less than $0.25$ thermodynamic
degrees of freedom for $U(2)$ in the perturbative center-symmetric
phase -- and $U(N>2)$ has even fewer~\cite{Schaden05}).

In these models confinement due to center-symmetry apparently is
quite separate from the existence of a mass gap. The latter might
arise due to a distinct (non-perturbative) mechanism. At weak
coupling, the correlations due to the non-trivial ground state
necessarily are long range and universality arguments that relate
the behavior near phase-transitions to spin models with
short-range interactions fail. Indeed, the phase transition at
\emph{weak coupling} is expected to be of first
order~\cite{Yaffe82} also in $SU(2)$-LGT. The extended
$SU(3)$-model at weak coupling shows two separate first order
phase transitions at temperatures $T_{c2} \sim 1.1\, T_c$. In the
intermediate phase the Polyakov loop attains only $1/3$ of its
maximal possible value.

The free energy to one loop was computed in the presence of
certain external sources. Insofar as these sources mimic higher
order perturbative and non-perturbative corrections, the observed
pattern of phase transitions perhaps is a qualitative possibility.
It thus is noteworthy that the Polyakov loop jumps to just $40\pm
10\%$ of its maximal possible value at the  deconfinement
transition in lattice simulations~\cite{Karsch00} and increases
rather slowly with the temperature thereafter. Heavy ion
experiments also do not observe a gas of free quarks and gluons
near the phase transition~\cite{Hirano05}.

That severe infrared divergences~\cite{Linde80} may prohibit a
fully broken perturbative ground state at high temperatures could
qualitatively explain some of these observations. The infrared
divergence would eliminate the option of a first order transition
to the fully broken perturbative vacuum.  One therefore might
expect that the corresponding phase transition moves to higher
temperatures and eventually disappears altogether in a calculation
of the free energy to higher \emph{perturbative} order.  In
$SU(2)$ the remaining possibility at weak coupling would be a
second order transition. In $U(3)$ (and presumably also $U(N>2)$ a
first order transition to a state with broken center-symmetry (but
non-maximal expectation of the Polyakov loop) would remain a
possibility. The pattern of symmetry breaking of the extended
models therefore thus could be more realistic if a transition to
the trivial vacuum can be excluded at weak coupling.
Phenomenologically, this (perturbative) restriction on the allowed
phase transitions can be simulated by a Sutherland potential for
the eigenvalues~\cite{Polychronakos99} or a (weaker) logarithmic
repulsion~\cite{Dumitru04}.

Recently an extreme version of the extended models was studied on
the lattice~\cite{Langfeld05}. The configuration space of an
SU(2)-LGT in this case was constrained to center-symmetric orbits
only. This constrained SU(2)-LGT reproduces the non-perturbative
$\beta$-function of full SU(2)-LGT at $T\sim 0$ but instead of a
deconfinement phase transition appears to have a
Hagedorn-transition~\cite{Hagedorn65}. The constrained model may
therefore be more amenable to a string
description~\cite{Pisarski82} than usual LGT.  On a periodic
lattice the constraint on the configuration space can be readily
implemented in a particular (ghost-free) Polyakov gauge for any
$SU(N)$-LGT. The transfer-matrix of the model (\equ{transf}) can
be formulated at any finite temperature. A Hamiltonian
construction of constrained models at large $N$ could be of
considerable interest, since they confine at weak
coupling~\cite{Schaden05,LargeNHag04}.

\noindent{\bf Acknowlegements:} Support for LC2005 from the CSSM
in Adelaide and Rutgers University is gratefully acknowledged. I
greatly enjoyed the hospitality of Adelaide University and
especially wish to thank L. v. Smekal, A. Williams, A. Kalloniatis
and D. Leinweber for numerous interesting discussions. I am
indebted to J. Rafaelski for explaining certain aspects of a
Hagedorn transition. 
\end{document}